# SBVR vs OCL: A Comparative Analysis of Standards

Imran Sarwar Bajwa, Behzad Bordbar, Mark Lee
School of Computer Science
University of Birmingham
Birmingham, UK
i.s.bajwa; b.bordbar; m.g.lee @cs.bham.ac.uk

*Abstract*— **In software modelling, the designers have to produce UML visual models with software constraints. Similarly, in business modelling, designers have to model business processes using business constraints (business rules). Constraints are the key components in the skeleton of business or software models. A designer has to write constraints to semantically compliment business models or UML models and finally implementing the constraints into business processes or source code. Business constraints/rules can be written using SBVR (Semantics of Business Vocabulary and Rules) while OCL (Object Constraint Language) is the well-known medium for writing software constraints. SBVR and OCL are two significant standards from OMG. Both standards are principally different as SBVR is typically used in business domains and OCL is employed to compliment software models. However, we have identified a few similarities in both standards that are interesting to study. In this paper, we have performed a comparative analysis of both standards as we are looking for a mechanism for automatic transformation of SBVR to OCL. The major emphasis of the study is to highlight principal features of SBVR and OCL such as similarities, differences and key parameters on which these both standards can work together.**

## I. INTRODUCTION

Software modeling is a major phase of software development aiming at curbing the complexity of the engineering of large software systems. UML (Unified Modeling Language) [1] is now affected as the defacto standard language for modelling of complex systems. However, models can be considerably accurate if constraints are included. The constraints clarify the semantics of UML models. OCL (Object Constraint Language), an OMG's standard, is a formal specification language [2] used to describe expressions and constraints on object-oriented (UML and MOF [5]) models. A constraint is a restriction on one or more values of (part of) an object-oriented model or system. OCL can not also be used as a declarative language for stating rules that apply to UML model but also appropriate for business use. OCL is a side-effect free language [2] which has mathematical foundation (first-order logic).

In business process modeling, the key phase is writing business rules. The Semantic of Business Vocabulary in Rules (SBVR) [3] is an adopted standard of the OMG and typically used to formalize complex compliance rule [3] (section: 12.1.2). Moreover, SBVR is an integral part of MDA (Model Driven Architecture) [4] as SBVR provides interoperable natural language models in MOF/XMI [5], [19] which bridges the gap between the world of thought and the technology world. SBVR makes use of natural languages to represent business rules by employing formal semantics. The formal semantics in SBVR are based on certain ceremonial approaches such as typed predicate logic, arithmetic, set and bag comprehension and modal logic [3]. Another significant feature is that SBVR is based on typed logic, First-Order-Logic (FOL) [3] (section: 2.2.6). SBVR also supports XMI schema for the interchange of business vocabularies and business rules among organizations and between software tools.

In this paper, we performed a comparative analysis of SBVR and OCL to identify their principal features such as similarities, differences and key parameters on which these both standards can work together. Apparently, SBVR and OCL are two different standards such as OCL is a medium for expressing constraints for UML models [2] and SBVR is used for specifying business rules for business processes [3]. However, there are a few similarities in both standards that are interesting to explore such as both standards are based on FOL [2], [3]. In this paper, we present in detail the results of the comparative analysis of SBVR and OCL. This analysis is used in transformation of SBVR rules to OCL constraints [31].

The rest of the paper is structured as follows. Section 2 highlights the criteria for comparing both standards. Section 3 describes a detailed one-to-one comparative analysis of SBVR and OCL. Section 4 presents discussion based on the comparison and the paper is closed with the conclusion section.

## II. CRITERIA FOR COMPARISON

We have identified seven major grounds on which we can perform a methodical comparison survey of SBVR and OCL standards. We identified these criteria on the basis of the features typically used to compare two formal languages such as OCL and B [10] and OCL and SQL [11].

*A. Relation to other Standards:* This section illustrates how SBVR and OCL relate to other allied standards and what is status of SBVR and OCL among other standards. Relationships will help in reducing gap among all these standards.

*B. Syntactical Features:* This section presents an account of all basic syntactic elements of SBVR that constitute a complete SBVR rule. The comparison of syntactic features of both SBVR and OCL will not only help in their model transformation but also their cross domain applications.

*C. Principal Features:* Principal features are basic or apparent characteristics of SBVR and OCL those are involved in making a simple SBVR rule and a common OCL constraint respectively. Moreover, we also explored strength an impact of SBVR an OCL in software and business modelling.

*D. Technical Features:* Technical features describe foundations and technical issues involved in SBVR and OCL standards. These issues help in understanding the technical gaps in SBVR and OCL and also hint on reducing these gaps.

*E. Expected Contributions:* This section explains how SBVR and OCL is contributing to business and IT communities. SBVR and OCL tools and their significance is also discussed in this section.

*F. Beneficiaries:* This section describes the beneficiary or recipients of both SBVR and OCL standards.

The major significance of the proposed criteria for comparison of SBVR and OCL that here both standards are not only compared with each other but also with other related standards i.e. criterion A. Criteria B, C, and D were defined specifically to support SBVR to OCL transformation. Criteria F and G states the contribution of both standards.

### III. SBVR VS OCL

This section presents one to one comparison of SBVR and OCL under the comparison criteria defined in the previous section. The details of the comparison are given below:

*A. Relation to other Standards*

*1. Model Driven Architecture (MDA):* SBVR is positioned to be entirely within the business model layer of the OMG's Model Driven Architecture (MDA) [4, Annex-A section: A.1]. SBVR is an integral part of MDA and SBVR's role in MDA has two dimensions. Primary dimension is a business design based on business rules and business vocabularies. Secondary dimension is a business model, including the models that SBVR supports, describe businesses and not the IT systems that support them. Overall SBVR enables the specific capture of terminology and meaning for any level of the MDA, so SBVR is used for PIM and PSM vocabularies and rules [3].

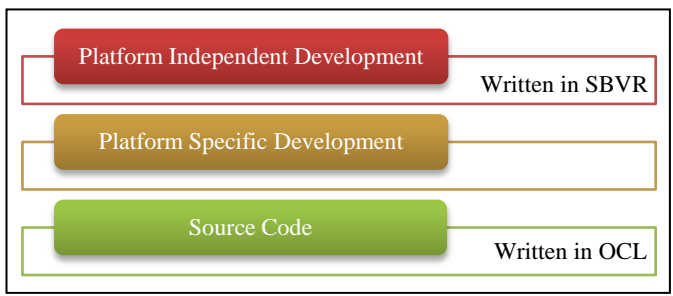

Figure 1. Role of OCL and SBVR in MDA

In MDA, OCL is a premier source for generating precise models and defining transformation definitions for models. Dedicated PIM and PSM [4] are typically written in UML & OCL. Moreover, in typical model transformation, OCL queries can be employed to depict elements of the source model and the elements of the target model of a transformation [20]. Moreover, OCL can also be utilized for definition of modeling languages.

*2. Model Object Facility (MOF):* SBVR provides interoperable natural language models in MOF/XMI) which bridges the gap between the world of thought and the technology world. SBVR principally uses OMG's MOF to provide interchange capabilities MOF/XMI mapping rules, enable generating MOF-compliant models and define an XML schema [3, Annex-A, Section: A.4.3]. Moreover, the SBVR Vocabulary is mapped to MOF elements that make up the SBVR Metamodel [3, section:13.2]. (see figure 5).

On other hand, a well-defined and named subset of OCL in the OCL specification is merely based on the common core of UML and MOF. [2] This compliance allows a subset of OCL to be used with both the MOF [21] and the UML. Moreover, OCL can be used at MOF layer and can help out in writing expressions of metamodels [2]. The primary usage of OCL with MOF has been in the definition of the UML metamodel. Several hundred of invariants also called "well-formed rules" are used to semantically complete a metamodel.

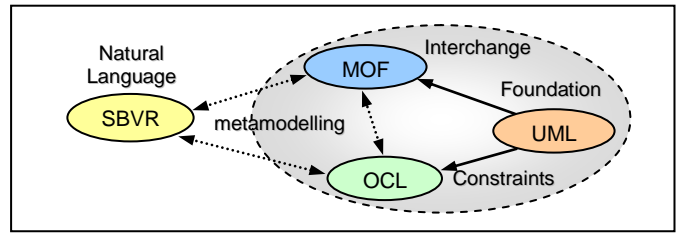

Figure 2. Relation of OCL and SBVR with MOF

*3. Unified Modeling Language (UML):* As, UML [1] is also based on MOF like SBVR as SBVR can be mapped to UML models [14] and back to SBVR [15]. The typical domain of SBVR is business modelling but SBVR based semantically formal representation can be used to capture software requirement specifications and improve automated software modelling process. Additionally, in SBVR, Internationalization [3, Annex-A, section:A.6.2] is proposed to handle the meanings of concepts (including fact types) and rules within a body of shared meanings expressed in different languages, both natural and artificial (e.g. UML and XML).

OCL is a part of the UML metamodel. OCL is typically used to specify constraints for UML models, while an OCL statement, without an accompanied UML model, refers to non-existing model elements; OCL can not use classes and associations.

*4. Business Process Management (BPM):* BPM's major focus is continuous optimization of business processes to achieve innovation, flexibility, and integration with technology in business models (see figure 6).

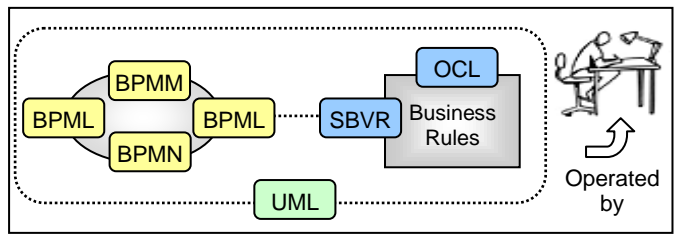

Figure 3. Status of OCL and SBVR in Business Process Management

In a complete process automation and management solution, business rules are the primary concern for defining process logic or process flow. Here comes the power of SBVR that is a prominent tool for defining the business rules. Moreover, a metamodel of business process can import SBVR's Vocabulary for describing Business Rules package in order to relate processes to rules.

However, OCL has no direct relation with BPM but OCL, due to its inherent declarative features, can be used for defining conditions and actions involved in a Business modelling. Moreover, OCL can be a very powerful way of stating business rules unambiguously as business rules are one of the pillar-post of BPM. Moreover, IBM has already presented business process modelling and simulation using UML that was passed on RUP (Rational Unified Process) [16]. RUP based UML can provide a systematic approach for visual representation of a business model. Here OCL will play an important role as a constraint language for the RUP based UML models.

*5. Service Oriented Architecture (SOA):* SOA helps in developing and managing flexible information system applications as well as to integrate the complex and assorted IT technologies. SBVR plays an important role at business process layer with BPMN and BPEL by recording business rules and modelling business processes. In addition to model business processes, SBVR based enterprise ontologies can be usedto define terms used by all the services deployed in the SOA. Moreover, SBVR can be helpful in middleware specifications and interface specifications used for CORBA services [3].

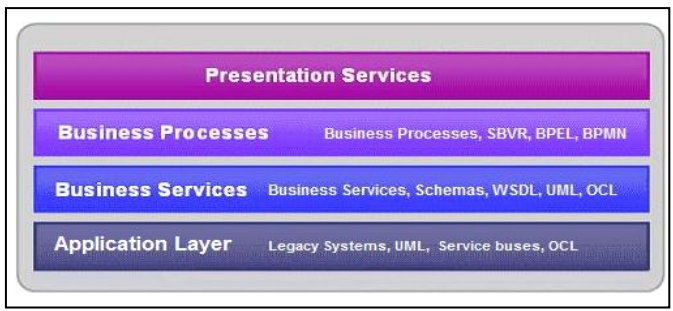

Figure 4. Role of OCL and SBVR in Service Oriented Architecture

On the other side, in SOA architecture, OCL is typically deployed at services layer with UML to record schemas and at application layer to represent application logic. Moreover, UML and OCL are typically used for specifying semantic web services compositions [17], [18].

*6. XML Metadata Interchange (XMI):* The XML Metadata Interchange (XMI) [19] is an OMG standard for exchanging metadata information via Extensible Markup Language (XML). SBVR is captured in terms of the MOF-based model created from the SBVR Vocabularies that includes the definitions of concepts, terms, business rules and other facts of the SBVR Metamodel. [3] (Annex-A, Section: A.5.3) MOF XMI is used as a notation for SBVR semantics as defined in SBVR Clauses 7-12.

On the other hand, OCL has similar relation to XMI as UML. [2] OCL can be used to represent models and metamodel with the help of XMI trees and OCL expressions can be exchanged using XMI. [2]

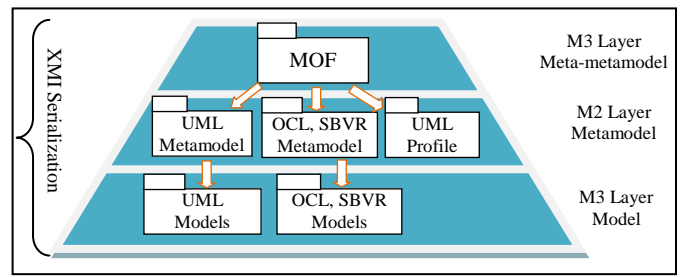

Figure 5. Relation of OCL and SBVR with XMI

*B. Syntactical Features*

*1. Vocabulary vs Classifiers:* A business vocabulary [3] contains all the specialized terms, concepts, and fact type forms of concepts that a given organization or community uses in their talking and writing in the course of doing business. SBVR vocabulary can be of two types: keywords and user defined elements. SBVR keywords are predefined terms that have fixed meanings (each, obligatory, necessary, at least, etc) and are adjuvant part of all SBVR rules..

On the other hand, similar to SBVR vocabularies, OCL expressions can refer to *Classifiers*, e.g., types, classes, interfaces, associations (acting as types), and data types. OCL keywords e.g. context, inv, pre, post, etc are the auxiliary elements of basic OCL expressions. Following is the comparison of major SBVR and OCL constituents:

*2. Noun Concept vs Context:* In SBVR metamodel, a Noun Concept [3] (section: 8.1.1) can be an object type [3] (section: 8.1.1) or an individual concept [3] (section: 8.1.1). Typically common nouns in English are classified as object types and proper nouns are classified as individual concepts.

In an OCL expression, *Context* [2, section: 7.3.5] is typically represented using a UML class. SBVR object type and individual concept can be equivalent to a context in an OCL expression.

*3. Verb Concepts vs Classifier AnyType*: In SBVR, the verb concepts [3] (section: 8.1.1) (action verbs) typically represent operations performed by/for a business entity. Action verbs in English can be matched to the method and operation names with out side-effect in OCL. The Verb *Concepts* (action verbs) in SBVR metamodel can be equivalent to classifier *AnyType* [2] (section: 8.2) in OCL metamodel. AnyType has a unique instance named *OclAny* and AnyType is defined to allow defining generic operations [2, section: 7.5.2) that can be invoked in any object or primitive literal value. [2]

Similarly, the class attributes [2, section: 7.5.1] without side-effects in UML also come under OCL's classifier *AnyType* [2, section: 8.2]. OCL attributes can be equivalent to SBVR's *Characteristics* [3, section: 11.1.2].

*4. Fact Types vs Associations*: Associations' ends [2, section: 7.5.3] are commonly used in OCL types. Similarly, in SBVR associations are supports by different types of fact types [3, section: 11.1.5.1] such as associations are represented using associative fact types [3, section: 11.1.5.1], aggregations are represented using categorization fact types [3, section: 11.1.5.1], and generalizations are represented using partitive fact types [3, section: 11.1.5.1]

In an OCL expression, the multiplicity of an association is also used. OCL's association multiplicity can be equal to SBVR's quantification such as universal quantification [3: 9.2.6) and non-universal quantification [3: 9.2.6) .

*5. Projections vs Collections:* A set of Projections [3] (section:9.3) are defined in SBVR to handle one or more than one variables. Similarly, OCL introduces Collections [2] (section:7.5.11) to provide support for managing multiple variables. The SBVR's Set Projection [3] (section:9.3) is equivalent to OCL's Set Collection [2] (section:7.5.11) and SBVR's Bag Projection [3] (section:9.3) is equivalent to OCL's Bag Collection [2] (section:7.5.11).

There is another type of collection such as *Sequence* [2, section: 7.5.11] that is not supported in SBVR. Similarly, there is another type of projection in SBVR such as *Closed Projection* [3, section: 9.3] that is not supported in OCL.

*6. Rules vs Expression*: The Rules [3, section: 12.1.2] in SBVR represent the specifications or the meanings of business constraints. The Similar to Rules in SBVR, there are Expression [2, section: 8.3] in OCL that make up a basic OCL constraint.

SBVR rules are of two basic types: structural rule and operative rule. Similarly, OCL expressions are also of two types: structural and behavioural constraints. Relation of SBVR rules with OCL expressions is explained below:

*7. Structural Rule vs Invariant:* The SBVR structural rules [3, section: 12.1.2] represent the structure of a business models and their underlying entities. The SBVR structural rules supplement definitions by using conditions and restrictions. Similar to SBVR structural rules, in OCL invariants are used to represent a structural constraint. OCL invariants typically specify strucyural information of UML models.

*8. Behavioural Rule vs Pre/Post Condition:* The behavioural rules [3, section: 12.1.2] govern the behaviour of business activities and operations. The behavioural or operative rules are ones that direct the activities involved in the business affairs. Akin to behavioural rules in SBVR, OCL's behavioural constraints such as pre/post conditions are particularly specified to handle behaviour of respective methods of classes and objects. The OCL pre/post conditions also specify state change

*C. Principal Features*

The principal features such as application, use, output, nature, etc of both languages (SBVR & OCL) have been compared in this section.

*1. Conceptual Modeling:* The primary focus of both languages (SBVR and OCL) is same i.e. conceptual modeling just their application domains are different such as SBVR is primarily used for business modeling (in combination with BPMN/BPEL), while OCL is used for software modeling (in combination with UML) and is employed for large scale object oriented models.

*2. Declarative Languages:* SBVR and OCL are both declarative language. SBVR rules should be expressed declaratively in natural-language sentences for the business audience [3]. Similarly, OCL support declaration of OCL constraints used for software models. However, SBVR is slightly different as it supports natural language e.g. English based declarative description of business information while, the OCL is a typed language and has its own formal syntax [2].

*3. Requirement Engineering:* Typically, requirement engineering is the important phase of software development, where requirements might be documented in various forms, such as natural-language documents, use cases, user stories, or process specifications. SBVR is used here to capture software/business requirements in natural languages (such as English) [3]. Contrary to SBVR, OCL is employed at later stages of software development such as graphical modelling (UML / SysML / BPMN). Here, OCL's duty is to ensure precise modelling [2]. However, non-functional requirements are typically constraints and represented in OCL.

*4. Side-Effect Free:* Both SBVR and OCL are side-effect free languages. SBVR based rules are side-effect free as all SBVR rules are distinct from any enforcement defined for it [3]. Similarly, OCL is a pure expression language and OCL constraints are side-effect free [2]. Hence, the side-effect free OCL expression cannot change anything in the model and the state of the system, even though an OCL expression can be used to specify a state change.

*5. Well-Formed Expression:* The SBVR business rules should be expressed in such a way that they can be validated for correctness by business people. Business rules should be expressed in such a way that they can be verified against each other for consistency [2]. Similarly, OCL expressions are strictly typed [2]. All the OCL constraints are type-checked and syntactically parsed to make it sure that the OCL expressions are well-formed expressions.

*D. Technical Features*

A set of technical features of both SBVR and OCL are compared in this section.

*1. SBVR is based on Formal Logics:* The formal semantics of SBVR is based on the following formal approaches: typed predicate logic; arithmetic; set and bag comprehension with some additional basic results from modal logic. [3, section: 10.1.2) The logic is essentially classical logic, so mapping to various logic-based languages is simple.

Similar to other formal specification languages, OCL also has its roots in mathematical logic. Although, OCL has mathematical foundation but no mathematical symbols are used. OCL is based on set theory and predicate logic and has a formal mathematical semantics. [30]

*2. Formal Semantics for SBVR:* A set of logical formulations have been defined in SBVR 1.0 document [3, section: 9.1] to provide a foundation for formal semantics. Typically, a business glossary or an enterprise vocabulary based information models are used by the business stakeholders for formal semantics. More formal semantics can be added through business facts and business rules. [3]

OCL constraints are also semantically formal as OCL formal semantics are described using UML. To specify the semantics of OCL expressions using semantic domain and the semantic domain is described in the form of a UML package, containing a UML class diagram, classes, associations, and attributes. [2, section: 10.1]

*3. Two-value Logic vs Three-value Logic:* SBVR's underlying logic is isomorphic (standard truth-functional logic) rather than epistemic logic. [3] Ultimately all ground facts are existential or elementary. The truth functional logic is two-valued, with negated existential formulae being used to avoid the use of null values.

Contrary to SBVR, OCL is based on a three-valued logic. OCL Boolean expression can result in true, false or undefined [2]. The three-valued logic can result in unexpected results.

*4. Inherent Extensibility:* An extended SBVR vocabulary is created by including the SBVR vocabulary into another business vocabulary that has other designations. The SBVR Vocabularies given by this specification are based on the English language, but can be used to define vocabularies in any language. Use of an SBVR vocabulary outside this specification does not change the SBVR vocabulary itself, but only uses it by way of reference.

Similarly, OCL inherits UML vocabulary (classes, associations, methods, etc) to complete basic OCL expressions.

### E. Expected Contributions

SBVR is relatively new standard than OCL. Yet some SBVR tools are available. Amit presented his work to transform SBVR business design to UML models [5]. He has used model driven engineering approach to transform SBVR specification into different UML diagrams e.g. activity diagram, sequence diagram, class diagram. Linehan's work [14] for the transformation of SBVR rules into formal languages e.g. Java is one of the only case. SBeaVeR is an [18] another open-source tool that translates the SBVR rules into Prolog rules. This also provides facility of expressing SBVR rule in "Structured English".

A few OCL tools have been presented for OCL parsing and type checking. IBM OCL Parser [23] was one of the first OCL tools written in Java by IBM. Dresden OCL Toolkit [24] is another OCL compiler. Similarly, USE (UML-based Specification Environment) tool [25] also presents an approach for the validation of UML models and OCL constraints. Other famous OCL tools are OCL-Toolkit [26], Cybernetic OCL Parser [27], ArgoUML [28], ModelRun [29], etc.

### F. Beneficiaries

SBVR is used the different groups of people who will benefit from it such as Business Analysts and Modelers, Business Vocabulary + Rules Integrators/ Administrators, Tool Builders, Logicians, Semanticists, and Linguists. Similar to SBVR OCL has a wide range of users e.g. OCL users are UML people, requirement people.

## IV. DISCUSSION

Though the SBVR and the OCL have some principal differences, yet there are some significant similarities in both standards. On the basis of comparison presented in section 4, following commonalties are identified in SBVR and OCL:

- Both are adopted standard of OMG.
- Both are declarative languages
- Both are used for defining constraints
- Both are side effects free.
- Both are integral part of the OMG's Model Driven Architecture (MDA).
- Both are based on mathematical logic i.e. FOL
- Both are not programming languages but the formal specification languages.
- Both support formal semantics
- Evaluation of the both (OCL/SBVR) languages' expression is instantaneous.
- Both can be used to specify structural and behavioural information.
- Both support messages and actions for a target.
- Both support *well-formed* rules.
- Both can be model transformed to each other.
- Both SBVR and OCL are formal languages, i.e., they remain easy to read and write.
- Both can be used as business modeling language.

OCL has been developed to fill the gap of traditional formal languages because they are usable to persons with a string mathematical background, but difficult for the average business or system modeler to use. Some major features explicitly exhibited by OCL and not supported by SBVR are following:

- OCL is standard query language and provides support to write queries but SBVR does not provide query support.
- OCL is not a programming language and it is not possible to write program logic or flow control in OCL.
- It is not possible to invoke processes or activate non-query operations within OCL because OCL is a modeling language in the first place, not everything in it is promised to be directly executable.
- In OCL, each OCL expression must be type conformant so it is not possible to compare an Integer with a String.
- As a modeling language, all implementation issues are out of scope and cannot be expressed in OCL.
- OCL uses OCL expressions and they are conceptually atomic. The state of the objects in the system cannot change during evaluation.

There are some features (see figure 6) those are explicitly exhibited by SBVR and not supported by OCL. In the result of the comparative study of SBVR and OCL, a set of interesting questions (see Table 1) came across those may be worthy to address for further technological advancements in the fields of both soft and business modeling.

Table 1. SBVR vs OCL

| SBVR | OCL |
|---|---|
| - SBVR is not query language<br>- SBVR does not support Sequence Collection.<br>- SBVR is typically designed for Business rules specification | - OCL is a query language<br>- OCL does not support Closed Projection.<br>- OCL does not support graphical notation<br>- OCL is designed for UML model constraints |

- Can OCL can be helpful in business modeling and if yes then how much?
- Can OCL constraints replace business rules in business process modeling as OCL can be model transformed to Java code?
- As, OCL can be used for specifying models for analysis purposes such as UML to Alloy transformation [28], can OCL be used for analysis of business processes?
- How SBVR can be useful in model transformations to RDF, OWL, Alloy, B, UML, BPMN, BPEL, ORM, etc?
- Can SBVR and OCL be used in combination as textual modeling languages?

## V. CONCLUSION

In this paper, we have investigated the use of a formal method to capture constraints in business and software domain models. We have compared the SBVR with OCL (together with its commercially-available tool support) with their syntactical, principal and technical features. We have also explored SBVR and OCL's relationship with other important standards such as MDA, MOF, UML, BPM, SOA, etc. The comparison shows a remarkable similarity between the two. The advantages in using SBVR and OCL are that both are mathematically based so that formal reasoning can be used to deduce desirable (and potentially undesirable) properties. Support for the SBVR and OCL is available via tools. However, the disadvantage with OCL is that there is lot of margin for improvement in terms of usability. However, the study shows that by using SBVR in requirement engineering and formal specification representation, we can attain beauty of natural languages and achieve power of formal languages.